# Strain Engineering of Quantum emitters in Hexagonal Boron Nitride

*Noah Mendelson, Marcus Doherty, Milos Toth, Igor Aharonovich, and Toan Trong Tran*

Mr. Noah Mendelson, Prof Milos Toth, Prof Igor Aharonovich, and Dr. Toan Trong Tran

School of Mathematical and Physical Sciences, University of Technology Sydney, Ultimo, New South Wales 2007, Australia.

Emails: igor.aharonovich@uts.edu.au; trongtoan.tran@uts.edu.au

Dr Marcus Doherty Laser Physics Centre, Research School of Physics, Australian National University, Australian Capital Territory 2601, Australia.

**Abstract:** Quantum emitters in hexagonal boron nitride (hBN) are promising building blocks for the realization of integrated quantum photonic systems. However, their spectral inhomogeneity currently limits their potential applications. Here, we apply tensile strain to quantum emitters embedded in few-layer hBN films and realize both red and blue spectral shifts with tuning magnitudes up to 65 meV, a record for any two-dimensional quantum source. We demonstrate reversible tuning of the emission and related photophysical properties. We also observe rotation of the optical dipole in response to strain, suggesting the presence of a second excited state. We derive a theoretical model to describe strain-based tuning in hBN, and the rotation of the optical dipole. Our work demonstrates the immense potential for strain tuning of quantum emitters in layered materials to enable their employment in scalable quantum photonic networks.

Single photon emitters (SPEs) embedded in solid state hosts are critical building blocks for a range of quantum technologies.[1-3] Integrating SPEs with on-chip nanophotonic components provides a scalable route towards the engineering of quantum gates and quantum circuitry.[4-6] However, unwanted interactions between the atom-like defects and the crystal host environment lead to spectral inhomogeneity that hinders device performance. To address this issue, methods for tuning emitter properties are critical for generating identical photons,[7-9] and for coupling to high-quality factor photonic resonators, where tuning magnitudes must be comparable to or greater than the cavity linewidths.[10]

Recently, hexagonal boron nitride (hBN), has been shown to host a range of sub-band gap defects operating as room temperature SPEs.[11-14] These SPEs display a number of desirable properties, including high photon purity,[15] bright emission,[16] and favorable quantum efficiencies.[17] However, the emitters have been shown to be susceptible to environmental influences, which lead to extreme inhomogeneity in their emission properties,[18] including a broad, continuous spectral range of zero phonon lines spanning from the deep ultraviolet to the near infrared.[19-21] Consequently, reliable tuning methods for controlling the emission properties are paramount for their implementation in quantum photonic applications.

Initial reports on tuning of hBN emitters employed voltage-controlled Stark shift devices and hydrostatic pressure.[22-25] Strain-based tuning of hBN defects has also been investigated using either the application of surface acoustic waves,[26, 27] or mechanical deflection of solid beams that translated vertical displacements to horizontal strain tensors.[28] In this work, we employ high degrees of tensile strain to tune the emission of hBN SPEs, and achieve record tuning magnitudes for a layered material of up to 20 nm (65meV). Unlike all previous reports, we take advantage of large area (~ few mm$^2$) ultrathin hBN films (~10 nm) that host a variety of SPEs.[29-32] These samples are amenable to the direct application of tensile strain (as is detailed below), and we report both red and blue spectral shifts, relating our results to modifications of the defect energy level manifold and corresponding coupling to the bulk phonon bath. We demonstrate a rotation of the optical dipole in select SPEs, suggesting the presence of a second excited state. A theoretical model to describe strain tuning the emission frequency of SPEs in hBN is fully derived and further expanded to conclusively confirm that dipole rotation occurs *via* the influence of this additional energy level. The ability to tune emission frequency and additional photophysical characteristics of emitters offers a promising route to tailor light-matter interactions in these systems[33].

Strain experiments were performed on hBN films grown by chemical vapor deposition (CVD) on a copper foil to a thickness of ~7 nm.[29] The films were transferred from the copper foil using a polymer-assisted (PMMA) wet-transfer process to a PDMS slab of 2.7 cm in length, and ~200 µm thick (*cf. methods*). The hBN/PDMS slab was secured in a mechanical straining device and mounted for optical characterization *via* confocal microscopy, as shown in **Figure** 1a. The PDMS slab was subject to varying degrees of tensile strain, as shown schematically in Figure 1b. Throughout the manuscript, we will discuss the strain applied to the PDMS substrate to interpret the corresponding results, expressed as the strain percentage *(S),* defined as $S(\%) = \frac{\Delta L}{L} * 100$, where L is the original length of the PDMS slab, and ΔL is the applied displacement. This provides an upper bound on the strain applied to the hBN film. Strain transfer from PDMS to 2D materials is low. However, the magnitude of transferred strain scales with the size of 2D sheet (greater strain transfer for larger flakes). In the case of large area CVD grown films can range from a few µm² to a few mm², giving a large variation in the transferred strain for different emitters[34]. As a result, the precise magnitude of strain transfer from PDMS to hBN varies for different emitters and is unresolved in our experiment. Incomplete strain transfer means the cited $S$ values of up to 7.4% may exaggerate the applied strain to hBN, which is unlikely to exceed a few percent.[35]

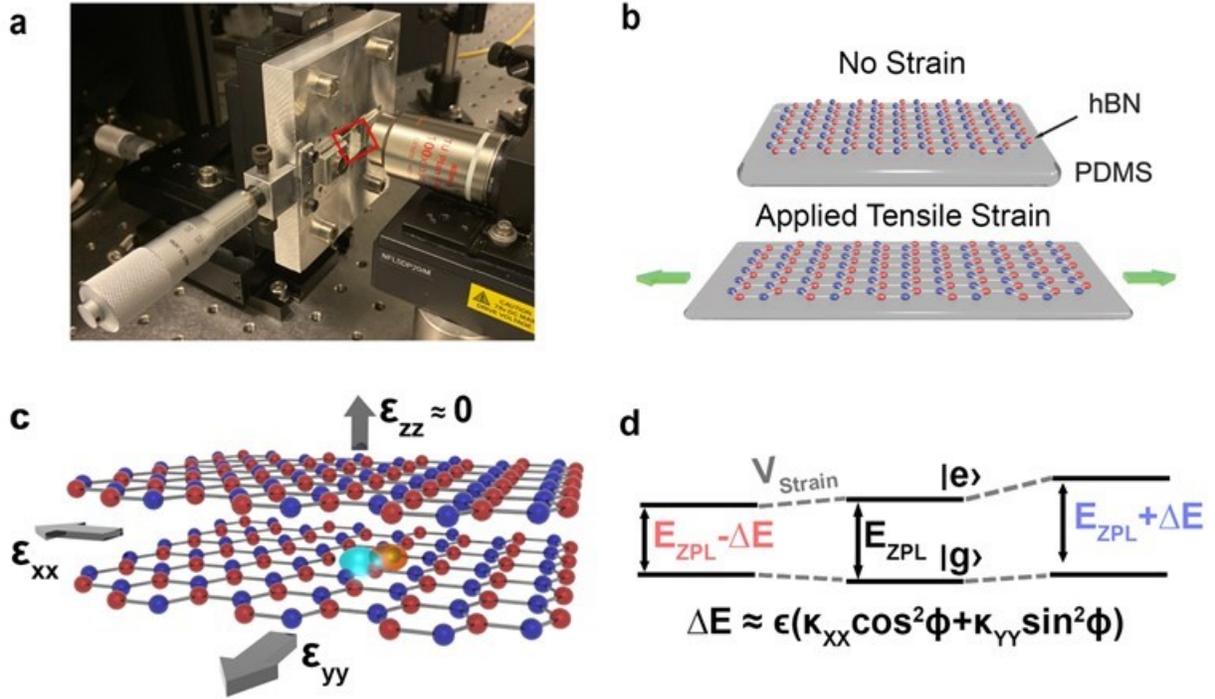

*Figure 1. Experimental setup and strain-induced blue shift of 13nm. a. Apparatus used to perform the strain tuning experiments. The red box highlights the PDMS slab with a ~7nm hBN film on top. b. Schematic showing the application of tensile strain to the PDMS slab. c. Schematic illustration of hBN film displaying a defect wavefunction. In our experiments strain is applied exclusively in the intra-layer domain. d. A simplified model showing the ground (|g>) and excited state (|e>) state levels for an atomic defect in hBN. When tensile strain is applied, $V_{Strain}$, the energy levels of the defect are modified, changing the emission energy of the defect ZPL by +/-ΔE according to equation 5.*

Emitters in hBN have been reported to exhibit a range of complex dependencies on the applied strain. For example, high-pressure measurements in which strain is applied isotropically have been interpreted in terms of a competition between the inter- and intra-layer strain tensors.[22] This, however, is not relevant to our experiment, where tensile strain is applied predominantly along the intralayer axis of hBN, meaning that the in-plane strain tensors will dominate the resulting optical response. The response of emitters to in-plane strain has been investigated by simulations of defect complexes of the type X-V, where X represents B, N, or heteroatom impurities, and V is a vacancy.[28] While the structural nature of the emitters remains unclear, recent simulations have confirmed X-V defects such as $N_BV_N$ and $C_BV_N$ are likely candidates, as also supported by our results.[36-38] Figure 1c depicts a simplified illustration of an atomic defect that acts as an SPE in hBN. Note, in our experiments, the orientation of the defect in relation to the crystal axes of hBN is not known.

First, we develop a theoretical model of strain interactions in hBN. We assume the defect to have $C_s$ symmetry. The electronic levels of defects observing $C_s$ symmetry can transform as one of two possible irreducible representations, A' and A''. For a specific defect, optical transitions may occur between two levels of the same representation (A'↔A' or A''↔A'') or two levels of different representation (A'↔A''). Ab-initio calculations of X-V type defects suggest that both the ground and first excited state transform as A' so here we consider the case of an A'↔A' optical transition. Importantly, the dipole moment of this transition must exist within the reflection of the defect. Applying group theory, the linear static strain interaction can be written as

$$V_{Strain} = (\sum_i \kappa_{ii} \epsilon_{ii} + 2\kappa_{xz}\epsilon_{xz})L_z + (\sum_i \kappa'_{ii} \epsilon_{ii} + 2\kappa'_{xz}\epsilon_{xz})L_x \tag{1}$$

The chosen coordinate system places z in the out-of-plane direction and x in the reflection plane of the defect (i.e. aligned with the optical dipole moment of the defect). The operators $L_z = \frac{1}{2}(|e\rangle\langle e| - |g\rangle\langle g|)$ and $L_x = \frac{1}{2}(|e\rangle\langle g| + |g\rangle\langle e|)$ are in the basis of the ground and excited states $\{|g\rangle, |e\rangle\}$ in their corresponding nuclear equilibrium configurations, while $\kappa_{ij}$ represent the linear strain susceptibility parameters, and $\epsilon_{ij}$ the strain components. For tensile strain applied along a single direction, the strain tensor components are

$$\epsilon_{ij} = \epsilon \cos\theta_i \cos\theta_j \qquad (2)$$

where $\epsilon = \frac{\Delta l}{l}$ is the strain magnitude and $\theta_i$ represents the angle between the axis of applied tensile strain and the $i^{th}$ coordinate direction. We note that in the following derivation, the small out-of-plane shear strain induced by the Poisson effect for a strain applied along a single crystal axis of the hBN flake (i.e, $\epsilon_{iz} \approx 0$ and $\epsilon_{zz} \approx 0$) is ignored. In which case the above simplifies to

$$V_{Strain} \approx (\kappa_{xx}\epsilon_{xx} + \kappa_{yy}\epsilon_{yy})L_z + (\kappa'_{xx}\epsilon_{xx} + \kappa'_{xx}\epsilon'_{xx})L_x = \epsilon(\kappa_{xx}\cos^2\theta_x + \kappa_{yy}\cos^2\theta_y)L_z + \epsilon(\kappa'_{xx}\cos^2\theta_x + \kappa'_{xx}\cos^2\theta_y)L_x \qquad (3)$$

Equation 3 can be rewritten in terms of $\phi$, the angle between strain axis and the defect dipole moment as

$$V_{Strain} = \epsilon(\kappa_{xx}\cos^2\phi + \kappa_{yy}\sin^2\phi)L_z + \epsilon(\kappa'_{xx}\cos^2\phi + \kappa'_{xx}\sin^2\phi)L_x \qquad (4)$$

The resulting first order change in the zero-phonon line (ZPL) energy is then

$$\Delta E \approx \epsilon\left(\kappa_{xx} \cos^2 \phi + \kappa_{yy} \sin^2 \phi\right) \qquad (5)$$

As a result, the linear strain susceptibility parameters $\kappa_{xx}$ and $\kappa_{yy}$ can be determined experimentally if the values $\epsilon$, $\phi$, and $\Delta E$ are resolved. Unfortunately, the uncertainty in the magnitude of strain transferred from PDMS to the hBN flake precludes an accurate determination of $\epsilon$ in the current work. However, this analysis will permit future studies to experimentally determine these values. The first order change in ZPL energy of a strained emitter is depicted schematically in Figure 1d in a simplified electron energy diagram where the initial ground state $|g\rangle$ and excited state $|e\rangle$ of a defect are modified upon the application of a strain field $V_{strain}$.

We use the above framework to explore the effects of tensile strain on the optical properties of SPEs in CVD-grown thin films of hBN. Figure 2a shows an SPE at 0% strain (red trace), with an initial ZPL position at 573.40 ± 0.08 nm, and an FWHM of 21.9 ± 0.8 nm. The inset is the corresponding second-order autocorrelation function, $g^2(\tau)$, confirming that the emitter is an SPE ($g^2(0)<0.5$). Upon application of 3.70% strain, the peak blue shifts to 560.82 ± 0.05 nm, a shift of ~12.6 nm (~ 49 meV). We note that this particular SPE had a dipole aligned with the strain field, i.e. $\phi \approx 0$. Using equation 5, where in this case the contribution of the $\kappa_{yy} \approx 0$ we can determine that the $\kappa_{xx}$ value for this example must be positive. Solid traces in Figure 2a display the Lorentzian fits for the ZPL at each position, and the extracted FWHMs are shown in Figure 2b. We observe a significant decrease in the broadening of the ZPL which narrows from 21.9 ± 0.8 nm to 11.7 ± 0.5 nm after the shift. This represents a nearly two-fold reduction in the ZPL linewidth (~9.0 nm), a process discussed in detail later.

Next, we investigate the observed shift of a second SPE upon applying tensile strain. Figure 2c shows PL spectra acquired *versus* time before straining (bottom panel), after the application of 5.55% strain (middle panel), and once the strain field is released (top panel). We observe a large and reversible red shift, and optical stability during the 100-second acquisition steps. The $g^2(\tau)$ and spectrum for each collection step are shown in figure S1. The fitted SPE peaks before (618.25 ± 0.09 nm) and after applying 5.55% strain (639.06 ± 0.06 nm) are plotted in figure 2d and demonstrate a red shift of ~ 20.8 nm (~ 65 meV). This constitutes the largest tuning magnitude for any 2D SPE to date. It is noted that the shoulder peak apparent in the unstrained emission spectrum is due to the fluorescence signal from PDMS (Figure S2). For this red shifted SPE we again find the dipole orientation to be nearly aligned with the strain field, figure S3, giving a value for $\phi \approx$ 8°. Analyzing equation 5 for this value of phi, we find the contribution of the $\kappa_{xx}$ term to be ~50x that of the $\kappa_{yy}$ term, suggesting that for this particular emitter $\kappa_{xx}$ is likely a negative value, in contrast to the previous emitter. There are a number of possible explanations for this discrepancy—that $\kappa_{xx}$ and $\kappa_{yy}$ may vary from emitter to emitter, that these values may differ by spectral location, or that the defects in question are not initially unstrained.

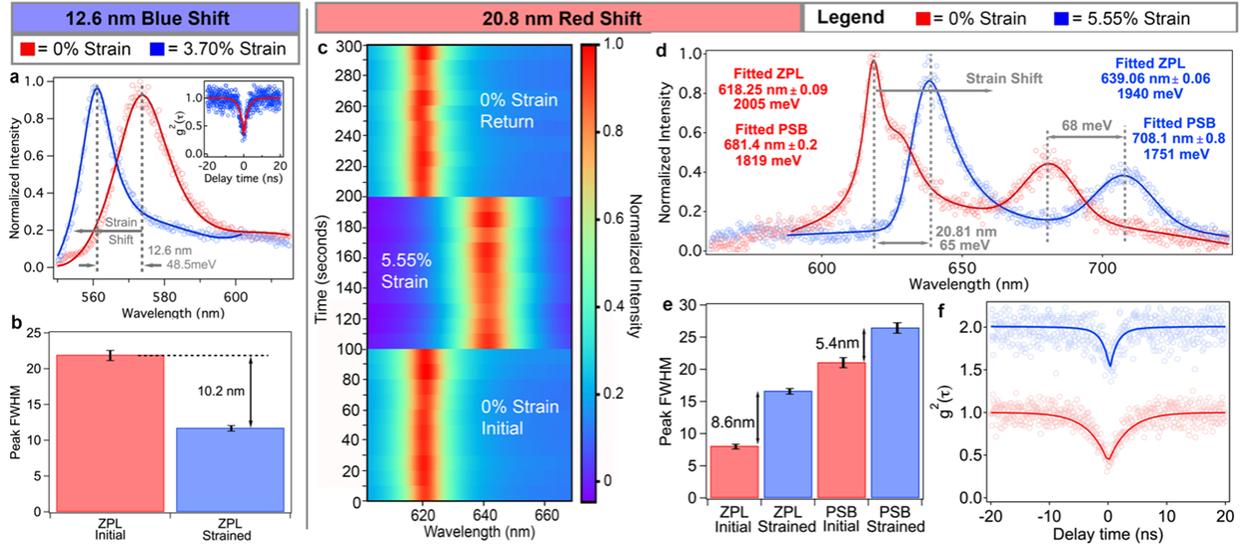

***Figure 2. Large red and blue strain-induced shifts.*** *In a-b, d-e red corresponds to initial SPEs while blue corresponds to shifted emission lines.* ***a.*** *A tensile strain of 3.70% is applied to an SPE initially at 573.40 ± 0.08 nm (red trace), inducing a blue shift of ~12.6 nm to 560.82 ± 0.04 nm (blue trace). Raw data for each (circles) is fitted with Lorentzian functions shown as a solid trace. Inset displays the $g^2(\tau)$ function confirming the quantum nature of the emission at 0% strain.* ***b.*** *The FWHM of both peaks are plotted showing a ~9.0 nm decrease in the associated linewidth of the emitter from 21.9 ± 0.8 nm (0% strain) to 12.9 ± 0.5 nm (3.7% strain).* ***c.*** *A time-resolved spectral acquisition for an SPE at 0% strain (bottom), 5.55% strain (middle), and returning to 0% strain (top). Each panel displays ten consecutive spectral acquisitions of 10 seconds each, demonstrating negligible spectral diffusion during the measurement, and the reversibility of the strain-induced red shift. There is a time offset between the three measurements, and each is normalized individually.* ***d.*** *Spectrum of the SPE at 0% strain (red trace) with a ZPL centered at 618.25 ± 0.09 nm, and at 5.55% strain (blue trace) with a ZPL centered at 639.06 ± 0.06 nm, showing a red shift of ~20.8 nm. Raw data for each (circles) is fitted with Lorentzian functions shown as a solid trace. See figure S2 regarding shoulder peak in 0% strain spectra.* ***e.*** *A bar graph plotting the FWHM of the ZPL and the PSB peaks at 0% and 5.55% strain, respectively. Both the*

*ZPL and PSB are significantly broadened upon straining. **f.** The $g^2(\tau)$ collection for the unstrained and strained SPE, showing a decrease of ~2.5ns for the extracted excited state lifetime of the emitter upon straining.*

Figure 2e plots the extracted FWHM of the ZPL and PSB before (red) and after (blue) applying 5.55% strain, where a broadening of 8.6 nm and 5.4 nm, respectively, are observed. For the ZPL peak this more than doubles the peak width from 8.1 ± 0.3 nm to 16.7 ± 0.4 nm, an increase of 106%, while the changes to the PSB are less prominent representing a FWHM increase of 25.5% from 21.1 ± 0.6 nm to 26.5 ± 0.8 nm. The red-shift induced increase in electron-phonon coupling lies in stark contrast to the decrease observed for the blue-shift in Figure 2a. In-fact all strain shifted SPEs follow this relationship between changes to the homogeneous broadening of the ZPL and PSB and the direction of the shift observed. Blue shifts decrease the prevalence of phonon coupling, while red-shifts increase the phonon coupling. We note that ZPL broadening accompanying red-shifts has also been reported for stark shifts of hBN emitters,[29] but has not previously been explained.

Below we focus on two potential explanations for this observation. Electron-phonon induced ZPL broadening can occur via two different mechanisms: first and second order acoustic phonon scattering[39]. First-order scattering is determined by the product of electron-phonon coupling and the density of phonon modes (ie Fermi's Golden Rule). We don't expect the density of phonon modes (ie phonon dispersion) to be significantly affected by strain. Thus, strain dependence of the ZPL width would be expected to result from modifying the electron-phonon coupling. As suggested in [37], this could arise from strain-induced modification of the defect's electronic wavefunction. Although, we can't identify another example of this occurring. Certainly, not in the context of quantum emitters in 3D solids. If this was the explanation, then these changes would need to correlate with the shift of the ZPL (ie blue shift is correlated with weaker coupling). Something that requires an ab initio theory survey in future work to properly assess.

Second-order Raman-type phonon scattering involving at least two electronic states (ie ground and first excited or first excited and a third state) is typically dominant at room temperature[40]. It also offers a more natural explanation for the correlation between ZPL shift and broadening through the introduction of a third factor into the rate expression that is inversely proportional to the square of the energy separation between the two electronic states involved in the scattering.[39] The observed blue shift with strain then correlates with an increase in this energy separation, and thus a reduction in the scattering rate a narrowing of the ZPL, as observed. The opposite also being true for the observed red shift with strain. Hence, we favor this second explanation as the simplest and most likely explanation.

We now turn our attention to the photophysical modifications of the emission upon applying strain. To this extent, we analyze the polarization of an emitter and record its photon statistics at each step. Figure 3 (a – d) shows the results of increasing tensile strain, at the values of 0% (red), 1.85% (green), 3.70% (light blue), and 5.55% (royal blue), where individual fits are displayed in figure S4. The red shifted ZPL peak moves 2.4 nm (~6 meV) at 5.5% strain, figure 3a. The $g^2(\tau)$ measurement at each strain value is shown in figure S5, demonstrating the quantum nature of the emission, and showing no changes to the SPE purity upon straining. Interestingly, the lifetime of the emitter stays approximately constant in this case. Consistent with investigations of higher energy defects (<690 nm), the red shift for this emitter is accompanied by a gradual broadening of the ZPL peak, suggesting an increased scattering between excited state energy levels, Figure 3b. Figure 3c displays the emission dipole angle as a function of the applied strain, showing a clear rotation upon straining. The dipole angle at 0% strain is ~ 158°, and gradually rotates with increasing strain to a value of ~ 144° at 5.55%, a ~ 14° rotation, moving towards the applied strain field at 90°(270°). Note, the hBN film is not rotated relative to the PDMS substrate, as evidenced by the confocal maps at each strain value (figure S6). To the best of our knowledge this is the first demonstration of a strain-induced dipole rotation for a room-temperature SPE.[41, 42]

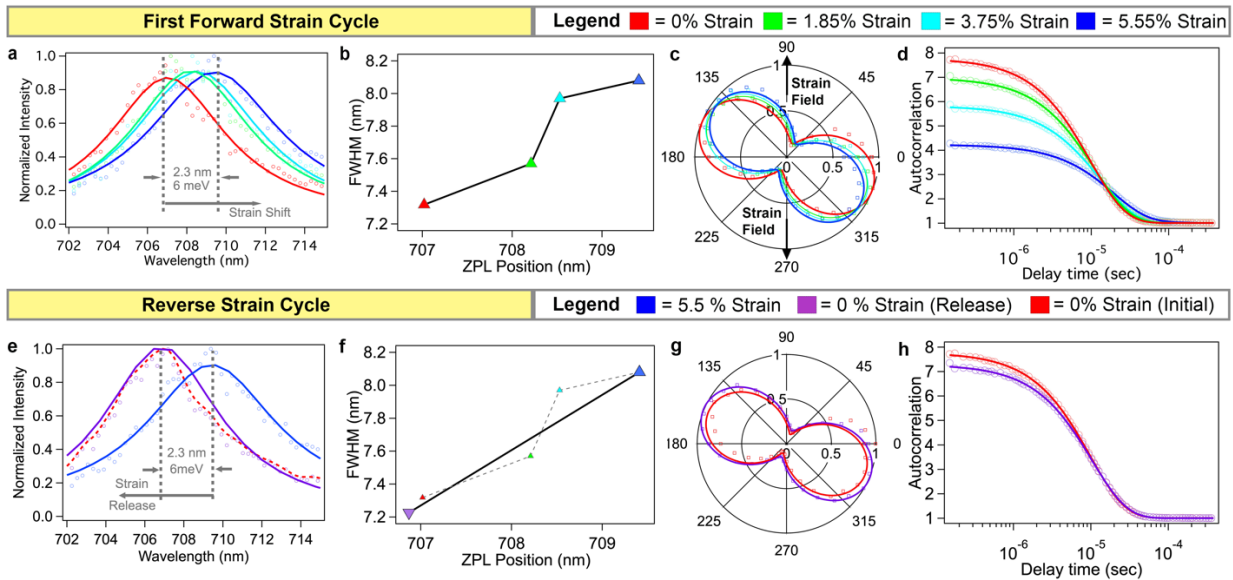

*Figure 3. Strain tuning the photophysical properties of hBN SPEs.* *The first row (a-d) displays a gradual increase in tensile strain from 0% to 5.55%. The second row (e-h) displays the reversing the strain from 5.55% back to 0%. The legend for both rows displays the strain percentage for a corresponding color.* ***a.*** *A gradual ZPL red shift of ~2.4 nm (~6meV) is observed for increasing strain values. Each collection is fit with a Lorentzian function (solid trace).* ***b.*** *The extracted ZPL FWHM vs position with increasing strain, displaying a gradual increase mirroring the observed spectral shift.* ***c.*** *Emission dipole orientation of the SPE with increasing strain. The tensile strain field is applied along the 90° (270°) axis, depicted with two arrows. A gradual rotation of the*

*dipole to a maximum displacement of ~12° is observed. **d.** Extended autocorrelation measurements for each strain %, fit with a double exponential function, suggesting the presence of two metastable states. The probability of transition into available meta-stable states is cut roughly in half at the maximum applied strain. **e.** Reversibility of the 2.4 nm red shift, while the red dashed line shows the initial spectrum for reference. Each collection is fit with Lorentzian functions (solid trace). **f.** The extracted ZPL FWHM vs position, displaying reversibility of the homogeneous broadening. Data points from the initial forward strain cycle are plotted in the background. **g.** Emission dipole orientation returns to its original position. **h.** Extended autocorrelation measurements, again fitted with a double exponential function, demonstrate the relative bunching is restored to initial values.*

The reorientation of the dipole in response to strain is an unexpected and a surprising result. The strain model provided above cannot describe such a reorientation without modification. To compensate for a strain induced dipole rotation, there must exist a third electronic level, one that transforms as A" and has the same spin multiplicity as the ground and first excited states. In such a case, the system can be represented as

$$V'_{Str} = (\sum_i \kappa''_{ii}\epsilon_{ii} + 2\kappa''_{xz}\epsilon_{xz})P_z + 2(\gamma_{xy}\epsilon_{xy} + \gamma_{yz}\epsilon_{yz})P_x + 2(\gamma'_{xy}\epsilon_{xy} + \gamma'_{yz}\epsilon_{yz})Q_x \quad (6)$$

where $P_z = \frac{1}{2}(|e2\rangle\langle e2| - |g\rangle\langle g|)$, $P_x = \frac{1}{2}(|e2\rangle\langle g| + |g\rangle\langle e2|)$, and $Q_x = \frac{1}{2}(|e2\rangle\langle e| + |e\rangle\langle e2|)$, where $|e2\rangle$ is the second excited state that transforms as A". The mixing between the second excited state and the other two states, governed by the second and third terms, is responsible for the dipole reorientation. It can be seen that this process occurs *via* a significant contribution of shear $\epsilon_{xy}$, and thus is only possible when the extension axis is not parallel with either the x or y

coordinates of the defect, i.e. when $\phi = 0°$ or $90°$. To further illustrate how the above interaction can cause a reorientation of the defect dipole moment, consider the special case where $\epsilon_{iz} = \epsilon_{zz} = \gamma_{xy} = \gamma_{yz} = \kappa'_{ii} = 0$. The strain interaction can then be written as

$$V_{Str} + V'_{Str} = \epsilon(\kappa_{xx} \cos^2 \phi + \kappa_{yy} \sin^2 \phi)L_z + \epsilon(\kappa''_{xx} \cos^2 \phi + \kappa''_{yy} \sin^2 \phi)P_z + 2(\gamma'_{xy} \cos \phi \sin \phi)Q_x \tag{7}$$

And the eigenstates for the ground, 1st, and 2nd excited states are approximately

$$|g(\epsilon)\rangle \approx |g\rangle \tag{8}$$

$$|e(\epsilon)\rangle \approx \cos\frac{\Omega}{2}|e\rangle - \sin\frac{\Omega}{2}|e2\rangle \tag{9}$$

$$|e2(\epsilon)\rangle \approx \cos\frac{\Omega}{2}|e2\rangle + \sin\frac{\Omega}{2}|e\rangle \tag{10}$$

where

$$\tan \Omega = \frac{2\epsilon(\gamma'_{xy} \cos \phi \sin \phi)}{\Delta E_{ex} + \epsilon\left((\kappa''_{xx} - \kappa_{xx}) \cos^2 \phi + (\kappa''_{yy} - \kappa_{yy}) \sin^2 \phi\right)} \tag{11}$$

and $\Delta E_{ex}$ is the zero-strain energy difference between $|e\rangle$ and $|e2\rangle$. Applying group theoretical selection rules, the optical dipole moment of the $|g(\epsilon)\rangle$ to $|e(\epsilon)\rangle$ transition is

$$\vec{d}(\epsilon) \approx d_1 \cos\frac{\Omega}{2}\hat{x} - d_2 \sin\frac{\Omega}{2}\hat{y} \qquad (12)$$

where $d_1 = |\langle g|\hat{d}|e\rangle|$ and $d_2 = |\langle g|\hat{d}|e2\rangle|$ are the dipole moments of the transitions to the two different excited states at zero strain. The new magnitude and orientation of the dipole moment are

$$d = \left(d_1^2 \cos^2\frac{\Omega}{2}\hat{x} + d_2^2 \sin^2\frac{\Omega}{2}\right)^{\frac{1}{2}} \qquad (13)$$

$$\Delta\phi(\epsilon) = \tan^{-1}\left(\frac{d_2}{d_1}\tan\frac{\Omega}{2}\right)$$

This demonstrates that the changes in the dipole moment depend on both the relative sizes of the zero-strain dipole moments of the two excited states and the strain induced mixing of the excited states. The dipole rotation observed shows a near linear trend when plotting the dipole orientation vs. the ZPL shift, figure S7. This is consistent with a first-order expansion of (13) in

$$\Delta\phi(\epsilon) \approx \frac{d_2}{d_1}\frac{\gamma'_{xy}\cos\phi\sin\phi}{\Delta E_{ex}}\epsilon \qquad (14)$$

Alignment between our experimental data and the corresponding theoretical model provides strong evidence for our conclusions of a second excited state. Confirmation of a third electronic state in this particular SPE also further indicates that the ZPL broadening trend is likely due to increased scattering between energy levels. These results do raise interesting questions of why the red shifted SPE in figure 2c-f, experiences a decrease in lifetime but no dipole rotation (figure S3), while the red shifted SPE in figure 3 displays dipole rotation however, no significant lifetime changes. The absence of dipole rotation in figure 2c-f may arise from the lack of shear strain ($\epsilon_{xy}$), as the optical dipole is nearly aligned with the extension direction ($\phi = 8°$) prior to straining. Note that we cannot eliminate the probability that the defects are simply of a different structural origin, however, we deem this scenario unlikely.

Figure 3d displays the second order extended auto-correlation measurements for the SPE as a function of applied strain. A significant and gradual decrease in long time photon bunching is observed as the strain field increases. Each curve is fitted with a double exponential, implying contributions from two available metastable states, according to equation S3. The degree of photon bunching observed is known to depend on the excitation power of the system,[43] as the population of the meta-stable state increases with the relative population of the excited state. In the current experiment, however, identical excitation conditions were used at each strain value. In a standard three level model, the probability of transition to a meta-stable state is governed by the rate coefficient $K_{23}$.[43] Long lived meta-stable states, such as those probed in figure 3d, can be the result of change in spin multiplicity (e.g. transition from a singlet to triplet state), transition to states which cannot be excited optically due to selection rules, or due to change in the charge state of the defect. The exact nature of the meta-stable states in hBN is not known, however, the data suggests that transition to the shorter lived meta stable state (with lifetimes on the order of a 1-10 µs) is significantly reduced, while the longer lived metastable state (with lifetimes on the order of 10-100 µs) experiences slightly increased population. The most straightforward explanation involves the changing position of the two excited states, evidenced by the dipole rotation, which modify the rate of intersystem crossing to the meta-stable states *via* changes in the spin-orbit coupling and the vibrational overlap of the available levels.

We next explore the reversibility of the strain-induced changes for the same SPE. Figure 3e shows the ZPL position is restored to its initial position (dashed red) upon releasing the applied strain (purple). Similarly, we find that the FWHM of the peak (Figure 3f), the dipole orientation (figure 3g), and the relative bunching (Figure 3h), are also restored to their initial values. Critically, this rules out the potential for layer slippage or permanent damage to the hBN lattice, such as ripping, confirming the strain-induced nature of the effects. Figure S8 displays a second forward strain cycle to higher maximum strain value of 7.40%, showing a larger red shift (~ 4.7 nm) and enhanced changes to the ZPL broadening, dipole rotation, and meta-stable state occupation rates.

Finally, we turn our attention to the overall trends observed across all investigated hBN SPEs. Figure 4a displays the relative change in the FWHM *versus* the shift magnitude and direction. Each circle represents a different SPE. In all recorded examples, the trend of ZPL narrowing with blue shifts and broadening with red shifts is conserved. At low shift magnitudes, the changes to the relative electron-phonon coupling appear to change faster, while the shift to peak width ratio changes drops off slightly for larger shift values.

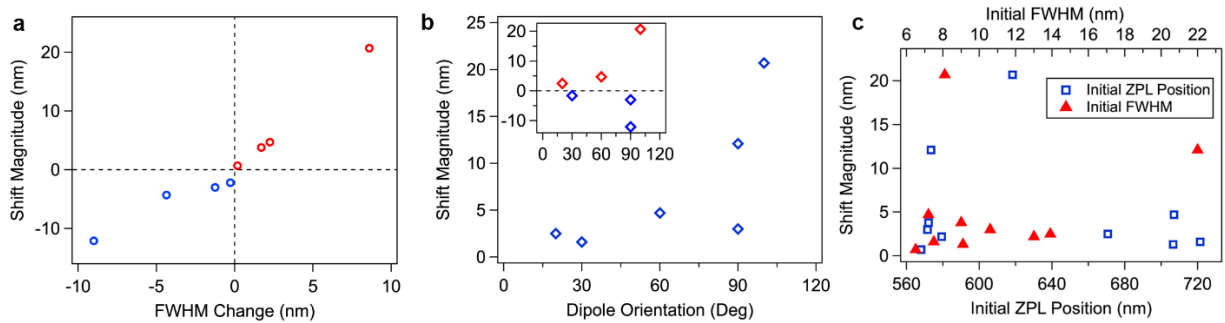

*Figure 4: Overall trends of strain modified quantum emission.* *Each panel displays a plot of the observed shift magnitude vs another independent variable. Each data point represents a different strained quantum emitter.* ***a.*** *vs change in the FWHM of the ZPL upon straining, showing each observed shift obeys the red/blue shift rule (i.e. blue shifts decrease e-phonon coupling, while red shifts increase this coupling).* ***b.*** *vs the dipole orientation of the unstrained SPE, showing a clear trend towards larger shift magnitudes when the dipole is aligned with the applied strain field.* ***c.***

*vs the initial ZPL position (0% strain), showing no observable dependence of shift on initial ZPL spectral position (red triangles). And vs initial FWHM (0% strain), showing no dependence of shift magnitude on the initial peak broadening (blue squares).*

Figure 4b displays the observed shift magnitudes as a function of the initial emission dipole angle, where the strain field is along the 90° axis. The inset in figure 4b shows that both large red and blue spectral shifts are observed for with near alignment between the optical dipole and the extension axis, further confirming that the sign and magnitude of $\kappa_{xx}$ changes between SPEs across the visible spectrum. Figure 4c, blue boxes, displays the observed spectral shifts as a function of the initial (0% strain) ZPL position. The majority of emitters exhibit a shift of up to ~ 5 nm, however, it was not uncommon to find extremely large spectral shifts exceeding 10 nm. We note again, that variation in the magnitudes of changes for any given SPE is correlated to the strain applied, $\epsilon$, which depends on unresolved features of the experiment, precluding direct determination of the strain applied, and accounting for the majority of variation in the parameters.

Figure 4c, red triangles, also plots the shift magnitude relative to the initial peak FWHM, showing no correlation between the magnitude of the spectral shifts and the homogeneous broadening of the SPE. This is an important observation, which suggests that similar shifts should persist even at cryogenic temperatures, where phonon broadening is reduced. This interpretation is supported by previous results showing up to ~ 9 nm shifts observed upon compressive strain applied through hydrostatic pressure at 20 K.[22] To contextualize the degree of shift magnitudes, we consider the largest strain-induced shift to be $10^5$ times greater than the expected natural linewidth of the emitter (~ 0.15 µeV linewidth vs ~ 65 meV spectral shift).

In summary, we have demonstrated record tuning magnitudes for a 2D quantum emitter of up to 65 meV at room temperature. The results were enabled by our newly established technique to grow ultra-thin hBN layers that can be easily transferred to a substrate of choice.[29] Importantly, such shift magnitudes help explain the broad and homogeneous distribution of hBN SPEs across the visible spectrum (550-800 nm), and effectively put a lower limit on the influence of strain at ~ 65 meV. We derived a model based of group theory to describe strain induced emission shifts in hBN. We determined the direction and magnitude of shifts are influenced by the alignment between the defect emission dipole and the applied strain field ($\phi$), the magnitude of strain applied $\epsilon$, and the strain susceptibility parameters $\kappa_{xx}$ and $\kappa_{yy}$. We found that blue (red) spectral shifts decrease (increase) the homogeneous broadening of the ZPL, likely through scattering between multiple electronic levels. Remarkably we demonstrated that the optical dipole of select SPEs can be rotated towards the applied strain field by up to 22°, showing a linear trend between dipole rotation and ZPL energy shift. We provide a detailed model predicting the linear dependence of dipole rotation, showing good agreement between theory and experiment, conclusively demonstrating the presence of an additional excited state. Finally, we confirmed that all changes are reversible, and the original photophysical properties of the SPEs are restored when the strain field is released.

Our work has several immediate implications in the field of integrated quantum photonics. First, such large shift magnitudes can eliminate hBN defects with inversion symmetry. Such defects are not expected to be amenable to high shifts under applied strain fields.[7] Therefore, the results support the assignation of the hBN defects in the visible range to defect of the $X_BV_N$ geometry (where X can be a nitrogen or carbon element). Second, we confirmed the presence of two excited states in at least some hBN quantum emitters, providing critical information to understand the level structure and behavior of these defects. Third, our results pave the way for future strain engineering of indistinguishable photons from hBN. Recently, this approach was fruitful to demonstrate indistinguishable photons from silicon-vacancy center (SiV) defects in diamond.[9, 44] We envision that two or more emitters in hBN can be put in resonance by employing the strain methods using cantilever geometries, as an example.[8, 45] The technique is fully amenable to the engineering of emitter – cavity coupling, whereby the SPE ZPL can be strain tuned into resonance with the cavity mode.[10, 46] Finally, and most intriguingly, the strain could be employed to enhance the optically detected magnetic resonance contrast.[47] Overall, our work constitutes a significant step forward in understanding light-matter interactions of quantum systems in 2D materials, in their leap towards scalable on-chip devices.

**Experimental Section:**

*hBN Growth and Transfer onto PDMS.* The hBN thin-films used in these experiments was fabricated via low-pressure chemical vapor deposition and transferred to the PDMS substrate following an established protocol.[29] Briefly, hBN was grown on copper, using ammonia borane as a precursor. Growth was performed at 1030 °C and a pressure of 2 Torr, in a 5% $H_2$/Ar atmosphere. The as-grown films were then transferred from copper to a PDMS polymer slab via a PMMA assisted wet transfer process. The polymer layer was then removed by soaking the sample in warm acetone (~50 °C) overnight, before further cleaning by exposure to UV-Ozone environment for 20 minutes.

*Optical Characterization.* PL studies were carried out using a home-built scanning confocal microscopy with continuous wave (CW) 532 nm laser (Gem 532, Laser Quantum Ltd.) as excitation. The laser was directed through a 532 nm line filter and a half-waveplate and focused onto the sample using a high numerical aperture (100×, NA = 0.9, Nikon) objective lens. Scanning was performed using an X−Y piezo fast steering mirror (FSM-300). The collected light was filtered using a 532 nm dichroic mirror (532 nm laser BrightLine, Semrock) and an additional long pass 568 nm filter (Semrock). The signal was then coupled into a graded-index multimode fiber, where the fiber aperture of 62.5 μm serves as a confocal pinhole. A flipping mirror was used to direct the emission to a spectrometer (Acton Spectra Pro, Princeton Instrument Inc.) or two avalanche photodiodes (Excelitas Technologies) in a Hanbury Brown- Twiss configuration, for collection of spectra and photon counting, respectively. Correlation measurements were carried out using a time-correlated single-photon counting module (PicoHarp 300, PicoQuant). All of the second-order autocorrelation $g^2(\tau)$ measurements were analyzed and fitted without background correction unless otherwise specified. For each SPE, ZPL and PSBs were fit with Lorentzian functions to extract both the peak centroid position and the FWHM of the peak.

**Supporting Information**
Supporting Information is available from the Wiley Online Library or from the author.

**Acknowledgments**

We thank Dr Carlo Bradac for fruitful discussions. The authors thank the Australian Research Council (DE170100169, DP180100070, DP190101058) and the Office of Naval Research Global under grant number N62909-18-1-2025 for financial support.

**References:**

1. Awschalom, D. D.; Hanson, R.; Wrachtrup, J.; Zhou, B. B., Quantum technologies with optically interfaced solid-state spins. *Nat. Photonics* **2018,** *12* (9), 516-527.

2. Atatüre, M.; Englund, D.; Vamivakas, N.; Lee, S.-Y.; Wrachtrup, J., Material platforms for spin-based photonic quantum technologies. *Nature Reviews Materials* **2018,** *3* (5), 38-51.

3. Aharonovich, I.; Englund, D.; Toth, M., Solid-state single-photon emitters. *Nat. Photonics* **2016,** *10* (10), 631-641.

4. Wang, J.; Sciarrino, F.; Laing, A.; Thompson, M. G., Integrated photonic quantum technologies. *Nature Photonics* **2019**.

5. Noel H. Wan, T.-J. L., Kevin C. Chen, Michael P. Walsh, Matthew E. Trusheim, Lorenzo De Santis, Eric A. Bersin, Isaac B. Harris, Sara L. Mouradian, Ian R. Christen, Edward S. Bielejec, Dirk Englund, Large-scale integration of near-indistinguishable artificial atoms in hybrid photonic circuits. *arXiv:1911.05265v1* **2019,** *arxiv.org*, e-Print archive https://arxiv.org/abs/1911.05265 (accessed Nov 18, 2019).

6. Nguyen, C. T.; Sukachev, D. D.; Bhaskar, M. K.; Machielse, B.; Levonian, D. S.; Knall, E. N.; Stroganov, P.; Riedinger, R.; Park, H.; Lončar, M.; Lukin, M. D., Quantum Network Nodes Based on Diamond Qubits with an Efficient Nanophotonic Interface. *Physical Review Letters* **2019,** *123* (18).

7. Meesala, S.; Sohn, Y.-I.; Pingault, B.; Shao, L.; Atikian, H. A.; Holzgrafe, J.; Gündoğan, M.; Stavrakas, C.; Sipahigil, A.; Chia, C.; Evans, R.; Burek, M. J.; Zhang, M.; Wu, L.; Pacheco, J. L.; Abraham, J.; Bielejec, E.; Lukin, M. D.; Atatüre, M.; Lončar, M., Strain engineering of the silicon-vacancy center in diamond. *Physical Review B* **2018,** *97* (20).

8. Sohn, Y. I.; Meesala, S.; Pingault, B.; Atikian, H. A.; Holzgrafe, J.; Gundogan, M.; Stavrakas, C.; Stanley, M. J.; Sipahigil, A.; Choi, J.; Zhang, M.; Pacheco, J. L.; Abraham, J.; Bielejec, E.; Lukin, M. D.; Atature, M.; Loncar, M., Controlling the coherence of a diamond spin qubit through its strain environment. *Nat Commun* **2018,** *9* (1), 2012.

9. Maity, S.; Shao, L.; Sohn, Y.-I.; Meesala, S.; Machielse, B.; Bielejec, E.; Markham, M.; Lončar, M., Spectral Alignment of Single-Photon Emitters in Diamond using Strain Gradient. *Physical Review Applied* **2018,** *10* (2).

10. Elshaari, A. W.; Buyukozer, E.; Zadeh, I. E.; Lettner, T.; Zhao, P.; Scholl, E.; Gyger, S.; Reimer, M. E.; Dalacu, D.; Poole, P. J.; Jons, K. D.; Zwiller, V., Strain-Tunable Quantum Integrated Photonics. *Nano Lett* **2018,** *18* (12), 7969-7976.

11. Konthasinghe, K.; Chakraborty, C.; Mathur, N.; Qiu, L.; Mukherjee, A.; Fuchs, G. D.; Vamivakas, A. N., Rabi oscillations and resonance fluorescence from a single hexagonal boron nitride quantum emitter. *Optica* **2019,** *6* (5).

12. Abidi, I. H.; Mendelson, N.; Tran, T. T.; Tyagi, A.; Zhuang, M.; Weng, L. T.; Özyilmaz, B.; Aharonovich, I.; Toth, M.; Luo, Z., Selective Defect Formation in Hexagonal Boron Nitride. *Advanced Optical Materials* **2019**.

13. Exarhos, A. L.; Hopper, D. A.; Patel, R. N.; Doherty, M. W.; Bassett, L. C., Magnetic-field-dependent quantum emission in hexagonal boron nitride at room temperature. *Nat Commun* **2019,** *10* (1), 222.

14. Bommer, A.; Becher, C., New insights into nonclassical light emission from defects in multi-layer hexagonal boron nitride. *Nanophotonics* **2019,** *0* (0).

15. Vogl, T.; Campbell, G.; Buchler, B. C.; Lu, Y.; Lam, P. K., Fabrication and Deterministic Transfer of High-Quality Quantum Emitters in Hexagonal Boron Nitride. *ACS Photonics* **2018,** *5* (6), 2305-2312.

16. Li, X.; Scully, R. A.; Shayan, K.; Luo, Y.; Strauf, S., Near-Unity Light Collection Efficiency from Quantum Emitters in Boron Nitride by Coupling to Metallo-Dielectric Antennas. *ACS Nano* **2019,** *13* (6), 6992-6997.

17. Nikolay, N.; Mendelson, N.; Özelci, E.; Sontheimer, B.; Böhm, F.; Kewes, G.; Toth, M.; Aharonovich, I.; Benson, O., Direct measurement of quantum efficiency of single-photon emitters in hexagonal boron nitride. *Optica* **2019,** *6* (8).

18. Li, C.; Xu, Z.-Q.; Mendelson, N.; Kianinia, M.; Toth, M.; Aharonovich, I., Purification of single-photon emission from hBN using post-processing treatments. *Nanophotonics* **2019,** *8* (11), 2049-2055.

19. Shevitski, B.; Gilbert, S. M.; Chen, C. T.; Kastl, C.; Barnard, E. S.; Wong, E.; Ogletree, D. F.; Watanabe, K.; Taniguchi, T.; Zettl, A.; Aloni, S., Blue-light-emitting color centers in high-quality hexagonal boron nitride. *Physical Review B* **2019,** *100* (15).

20. Jungwirth, N. R.; Calderon, B.; Ji, Y.; Spencer, M. G.; Flatte, M. E.; Fuchs, G. D., Temperature Dependence of Wavelength Selectable Zero-Phonon Emission from Single Defects in Hexagonal Boron Nitride. *Nano Lett* **2016,** *16* (10), 6052-6057.

21. Tran, T. T.; Elbadawi, C.; Totonjian, D.; Lobo, C. J.; Grosso, G.; Moon, H.; Englund, D. R.; Ford, M. J.; Aharonovich, I.; Toth, M., Robust Multicolor Single Photon Emission from Point Defects in Hexagonal Boron Nitride. *ACS Nano* **2016,** *10* (8), 7331-8.

22. Xue, Y.; Wang, H.; Tan, Q.; Zhang, J.; Yu, T.; Ding, K.; Jiang, D.; Dou, X.; Shi, J. J.; Sun, B. Q., Anomalous Pressure Characteristics of Defects in Hexagonal Boron Nitride Flakes. *ACS Nano* **2018,** *12* (7), 7127-7133.

23. Noh, G.; Choi, D.; Kim, J. H.; Im, D. G.; Kim, Y. H.; Seo, H.; Lee, J., Stark Tuning of Single-Photon Emitters in Hexagonal Boron Nitride. *Nano Lett* **2018,** *18* (8), 4710-4715.

24. Nikolay, N.; Mendelson, N.; Sadzak, N.; Böhm, F.; Tran, T. T.; Sontheimer, B.; Aharonovich, I.; Benson, O., Very Large and Reversible Stark-Shift Tuning of Single Emitters in Layered Hexagonal Boron Nitride. *Physical Review Applied* **2019,** *11* (4).

25. Scavuzzo, A.; Mangel, S.; Park, J.-H.; Lee, S.; Loc Duong, D.; Strelow, C.; Mews, A.; Burghard, M.; Kern, K., Electrically tunable quantum emitters in an ultrathin graphene–hexagonal boron nitride van der Waals heterostructure. *Applied Physics Letters* **2019,** *114* (6).

26. Lazić, S.; Espinha, A.; Pinilla Yanguas, S.; Gibaja, C.; Zamora, F.; Ares, P.; Chhowalla, M.; Paz, W. S.; Burgos, J. J. P.; Hernández-Mínguez, A.; Santos, P. V.; van der Meulen, H. P., Dynamically tuned non-classical light emission from atomic defects in hexagonal boron nitride. *Communications Physics* **2019,** *2* (1).

27. Iikawa, F.; Hernández-Mínguez, A.; Aharonovich, I.; Nakhaie, S.; Liou, Y.-T.; Lopes, J. M. J.; Santos, P. V., Acoustically modulated optical emission of hexagonal boron nitride layers. *Applied Physics Letters* **2019,** *114* (17).

28. Grosso, G.; Moon, H.; Lienhard, B.; Ali, S.; Efetov, D. K.; Furchi, M. M.; Jarillo-Herrero, P.; Ford, M. J.; Aharonovich, I.; Englund, D., Tunable and high-purity room temperature single-photon emission from atomic defects in hexagonal boron nitride. *Nat Commun* **2017,** *8* (1), 705.

29. Mendelson, N.; Xu, Z. Q.; Tran, T. T.; Kianinia, M.; Scott, J.; Bradac, C.; Aharonovich, I.; Toth, M., Engineering and Tuning of Quantum Emitters in Few-Layer Hexagonal Boron Nitride. *ACS Nano* **2019,** *13* (3), 3132-3140.

30. Comtet, J.; Glushkov, E.; Navikas, V.; Feng, J.; Babenko, V.; Hofmann, S.; Watanabe, K.; Taniguchi, T.; Radenovic, A., Wide-Field Spectral Super-Resolution Mapping of Optically Active Defects in Hexagonal Boron Nitride. *Nano Lett* **2019,** *19* (4), 2516-2523.

31. Song, L.; Ci, L.; Lu, H.; Sorokin, P. B.; Jin, C.; Ni, J.; Kvashnin, A. G.; Kvashnin, D. G.; Lou, J.; Yakobson, B. I.; Ajayan, P. M., Large scale growth and characterization of atomic hexagonal boron nitride layers. *Nano Lett* **2010,** *10* (8), 3209-15.

32. Stern, H. L.; Wang, R.; Fan, Y.; Mizuta, R.; Stewart, J. C.; Needham, L. M.; Roberts, T. D.; Wai, R.; Ginsberg, N. S.; Klenerman, D.; Hofmann, S.; Lee, S. F., Spectrally Resolved Photodynamics of Individual Emitters in Large-Area Monolayers of Hexagonal Boron Nitride. *ACS Nano* **2019,** *13* (4), 4538-4547.

33. Dai, Z.; Liu, L.; Zhang, Z., Strain Engineering of 2D Materials: Issues and Opportunities at the Interface. **2019,** *31* (45), 1805417.

34. Frisenda, R.; Drüppel, M.; Schmidt, R.; Michaelis de Vasconcellos, S.; Perez de Lara, D.; Bratschitsch, R.; Rohlfing, M.; Castellanos-Gomez, A., Biaxial strain tuning of the optical properties of single-layer transition metal dichalcogenides. *npj 2D Materials and Applications* **2017,** *1* (1), 10.

35. Gant, P.; Huang, P.; Pérez de Lara, D.; Guo, D.; Frisenda, R.; Castellanos-Gomez, A., A strain tunable single-layer MoS2 photodetector. *Materials Today* **2019,** *27*, 8-13.

36. Tawfik, S. A.; Ali, S.; Fronzi, M.; Kianinia, M.; Tran, T. T.; Stampfl, C.; Aharonovich, I.; Toth, M.; Ford, M. J., First-principles investigation of quantum emission from hBN defects. *Nanoscale* **2017,** *9* (36), 13575-13582.

37. Wu, F.; Galatas, A.; Sundararaman, R.; Rocca, D.; Ping, Y., First-principles engineering of charged defects for two-dimensional quantum technologies. *Physical Review Materials* **2017,** *1* (7).

38. Grosso, G.; Moon, H.; Lienhard, B.; Ali, S.; Efetov, D. K.; Furchi, M. M.; Jarillo-Herrero, P.; Ford, M. J.; Aharonovich, I.; Englund, D., Tunable and high-purity room temperature single-photon emission from atomic defects in hexagonal boron nitride. *Nat. Commun.* **2017,** *8* (1), 705.

39. Stoneham, A. M., *Theory of Defects in Solids*. Oxford University Press: 1975.

40. Fu, K. M. C.; Santori, C.; Barclay, P. E.; Rogers, L. J.; Manson, N. B.; Beausoleil, R. G., Observation of the Dynamic Jahn-Teller Effect in the Excited States of Nitrogen-Vacancy Centers in Diamond. *Phys. Rev. Lett.* **2009,** *103* (25), 256404.

41. Plumhof, J. D.; Křápek, V.; Ding, F.; Jöns, K. D.; Hafenbrak, R.; Klenovský, P.; Herklotz, A.; Dörr, K.; Michler, P.; Rastelli, A.; Schmidt, O. G., Strain-induced anticrossing of bright exciton levels in single self-assembled GaAs/AlxGa1−xAs and InxGa1−xAs/GaAs quantum dots. *Physical Review B* **2011,** *83* (12).

42. Martín-Sánchez, J.; Trotta, R.; Mariscal, A.; Serna, R.; Piredda, G.; Stroj, S.; Edlinger, J.; Schimpf, C.; Aberl, J.; Lettner, T.; Wildmann, J.; Huang, H.; Yuan, X.; Ziss, D.; Stangl, J.; Rastelli, A., Strain-tuning of the optical properties of semiconductor nanomaterials by integration onto piezoelectric actuators. *Semiconductor Science and Technology* **2018,** *33* (1).

43. Neu, E.; Agio, M.; Becher, C., Photophysics of single silicon vacancy centers in diamond: implications for single photon emission. *Opt Express* **2012,** *20* (18), 19956-71.

44. Evans, R. E.; Bhaskar, M. K.; Sukachev, D. D.; Nguyen, C. T.; Sipahigil, A.; Burek, M. J.; Machielse, B.; Zhang, G. H.; Zibrov, A. S.; Bielejec, E.; Park, H.; Lončar, M.; Lukin, M. D., Photon-mediated interactions between quantum emitters in a diamond nanocavity. *Science* **2018,** *362* (6415), 662-665.

45. Kim, H.; Moon, J. S.; Noh, G.; Lee, J.; Kim, J. H., Position and Frequency Control of Strain-Induced Quantum Emitters in WSe2 Monolayers. *Nano Lett* **2019**.

46. Sipahigil, A.; Evans, R. E.; Sukachev, D. D.; Burek, M. J.; Borregaard, J.; Bhaskar, M. K.; Nguyen, C. T.; Pacheco, J. L.; Atikian, H. A.; Meuwly, C.; Camacho, R. M.; Jelezko, F.; Bielejec, E.; Park, H.; Loncar, M.; Lukin, M. D., An integrated diamond nanophotonics platform for quantum-optical networks. *Science* **2016,** *354* (6314), 847-850.

47. Pingault, B.; Jarausch, D. D.; Hepp, C.; Klintberg, L.; Becker, J. N.; Markham, M.; Becher, C.; Atature, M., Coherent control of the silicon-vacancy spin in diamond. *Nat Commun* **2017,** *8*, 15579.